\documentclass[letterpaper]{article} 
\usepackage[preprint]{aaai2027}  
\usepackage[hyphens]{url}  
\usepackage{graphicx} 
\usepackage{natbib}  
\usepackage{caption} 
\usepackage{booktabs}
\usepackage{amsmath}
\usepackage{amssymb}

\title{Is Personalized Modality Weighting Actually Personalized? A Controlled Audit of Per-User Weighting Claims in Multimodal Recommenders}

\author{
    Jingyuan Zheng\textsuperscript{1},\quad
    Xin Zhang\textsuperscript{1},\quad
    Yang Gu\textsuperscript{2},\quad
    Dongjing Wang\textsuperscript{1},\quad
    Yuxiang Wang\textsuperscript{1},\\
    Xudong Shen\textsuperscript{3},\quad
    Haiping Zhang\textsuperscript{1},\quad
    Youhuizi Li\textsuperscript{1},\quad
    Dongjin Yu\textsuperscript{1}
}
\affiliations{
    \textsuperscript{1}School of Computer Science and Technology, Hangzhou Dianzi University, Hangzhou, China\\
    \textsuperscript{2}College of Humanities, Hangzhou Normal University, Hangzhou, China\\
    \textsuperscript{3}Fuxi AI Lab, NetEase Games, NetEase Inc., Hangzhou, China\\
    \texttt{zhengjoy, zhangxin, dongjing.wang, lsswyx, zhanghp, huizi, yudj\,@hdu.edu.cn}\quad
    \texttt{3073572279@qq.com}\quad
    \texttt{hzshenxudong@corp.netease.com}
}

\begin{document}

\maketitle

\begin{abstract}
Per-user modality weighting is deployed at billion-user scale in multimodal recommenders, through user modality-strength vectors, attention gates, meta-weight hypernetworks, and low-rank guided weights, each claiming a ranking gain from user-specific modality preference.
Yet, to our knowledge, prior evaluations do not isolate a genuinely user-specific signal from a global modality weight plus model capacity.
We audit this family with a two-contrast audit principle, reducing six implementations onto one shared collaborative backbone and measuring a utility gap (real-GM) against a single global modality weight and an identifiability gap (real-shuf) against an eval-time permutation of the user-weight binding.
Across three independent short-video corpora, a single global weight already delivers nearly all of the content gain (+1.9/+3.6/+3.5pp over a no-modality baseline, $p<.001$).
Making the weight per-user adds no consistent utility: no implementation wins on all corpora and metrics, and the few positive gaps are small ($\leq$0.9pp) and flip.
The shuffle control is necessary but not sufficient, since real-shuf reaches +128\% of the content gain for heads that simultaneously \emph{lose} to the global weight.
We trace this dissociation to gates reading the shared collaborative embedding: decoupling the gate input collapses the inflated real-shuf to near zero while the utility conclusion stands.
A monotone signal-implant dose-response (capture AUROC rising from 0.57 to 0.89 and from 0.64 to 1.00) verifies the harness would detect user-specific structure if present, and every finding replicates on a fourth, cross-domain e-commerce corpus.
We propose reporting real-GM alongside real-shuf as a minimum evidentiary standard for personalization claims.
\end{abstract}


\section{Introduction}

Multimodal recommender systems increasingly deploy \emph{per-user} modality weights, on the claim that users differ in how much they trust visual, acoustic, or textual content.
A billion-user framework learns a user modality-strength vector so that each profile scales the visual, textual, and acoustic channels on its own~\cite{chen2024m3csr}, and reliability-guided variants tie the weight to per-user modality-reliability estimates~\cite{dong2025margo}.
The same idea recurs as bilinear attention gates, meta-weight hypernetworks~\cite{shu2019metaweightnet}, and low-rank guided user weights, and each reports a ranking gain over an unweighted or globally weighted model.

Audit studies in collaborative filtering have found that complex methods often fail to
outperform well-tuned simple baselines once evaluation is made
fair~\cite{ferraridacrema2019progress,ferraridacrema2021troubling}, and the same pattern
appears in information retrieval~\cite{lin2019neuralhype}.
To our knowledge, prior per-user modality-weighting evaluations do not isolate a genuinely user-specific signal from a single global modality weight plus model capacity.
Standard ablations remove a modality or a module, but they do not directly break the binding between a user and the weight it receives, so a gain that a single global weight would also produce still counts as evidence for personalization.
Two confounds are left uncontrolled: a global modality scale that helps every user equally, and the extra capacity a per-user head adds to the backbone.

\begin{figure*}[t]
\centering
\includegraphics[width=\textwidth]{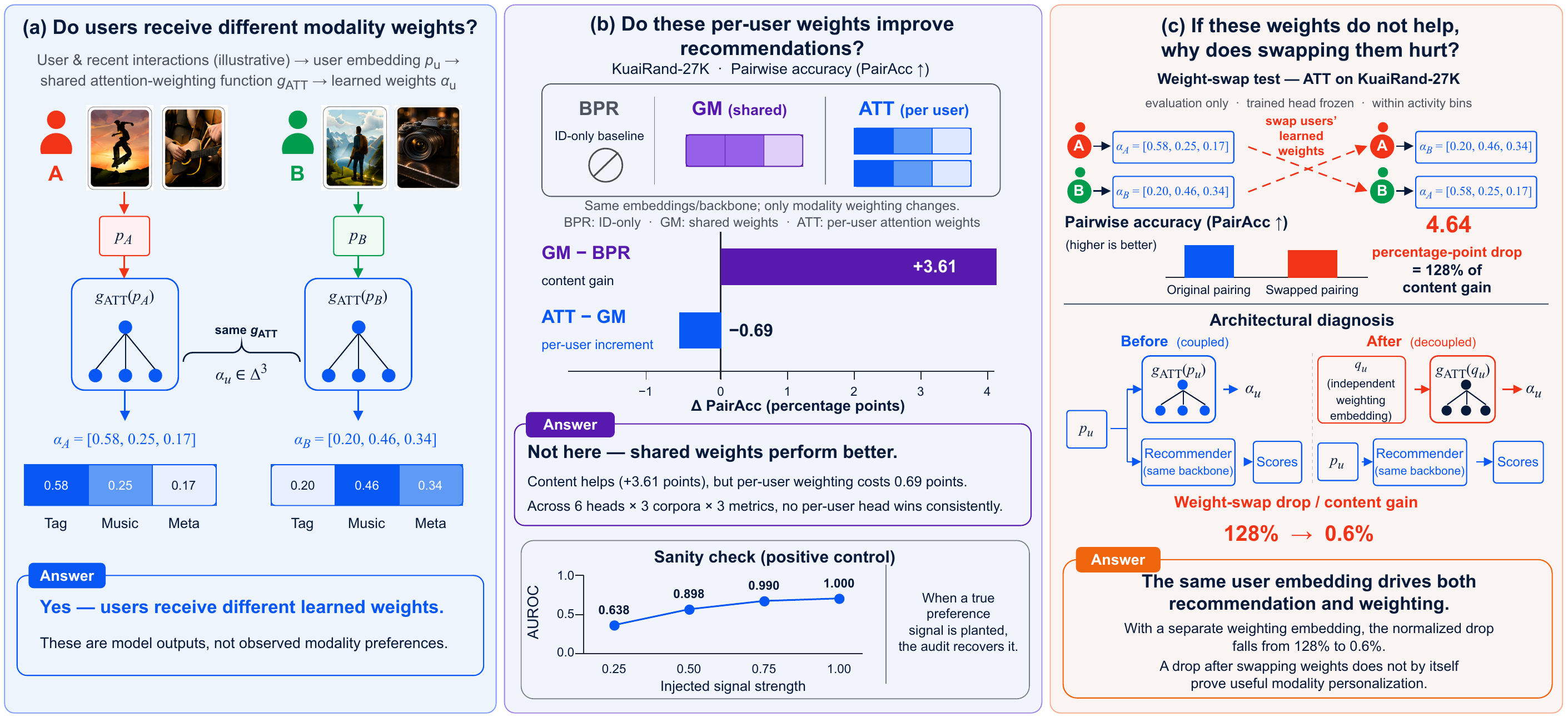}
\caption{Different per-user modality weights do not imply useful modality personalization. The three panels ask in turn whether users receive different weights (a), whether those weights beat a single global weight (b), and why swapping them still hurts (c). On KuaiRand-27K the attention head trails the global weight by 0.69pp (utility gap real-GM), yet an eval-time weight swap costs +4.64pp, equal to 128\% of the content gain (identifiability gap real-shuf). Decoupling the gate input collapses that swap effect to 0.6\%, so it is architectural, and a swap effect alone cannot certify useful personalization.}
\label{fig:teaser}
\end{figure*}

We therefore audit this family with two paired contrasts on one shared collaborative backbone: a \emph{utility gap} (real-GM) against a single global weight, and an \emph{identifiability gap} (real-shuf) against a frozen permutation of the user-weight binding (Figure~\ref{fig:teaser}).
Six weighting heads, from a free per-user table to attention gates, a hypernetwork, and a low-rank guided weight, are reduced onto the same backbone so that only the weight-production mechanism differs.
The audit covers this explicit weighting family, where personalization enters as a per-user scalar, vector, or gate over modality channels, rather than multimodal recommendation architectures at large.
A head is genuinely personalized only when it beats the global weight on the utility gap, not merely when it beats its own shuffle on the identifiability gap.

This audit makes three contributions.
\begin{itemize}
\item We formulate a general audit principle for personalization claims: a per-user mechanism is credible only when it beats the strongest non-personalized alternative on a utility contrast, not merely its own shuffle on an identifiability contrast. Instantiating it across six implementations and four corpora, no head beats a single global weight on all corpora across all metrics, and the global weight alone captures nearly all of the content gain (Audit Design; Results; Appendix).
\item Shuffle controls are necessary but not sufficient, because architectural confounding inflates the identifiability gap to +128\% of the content gain while utility stays negative, and decoupling the gate input removes the inflation causally (Results).
\item We release a two-contrast audit protocol with implant-based sensitivity calibration, five-seed paired tests with Benjamini--Hochberg FDR correction, and all per-seed results, a statistical standard not previously applied to this family of claims (Audit Design).
\end{itemize}

Our sharpest finding is a dissociation, where heads that look strongly personalized under a shuffle control simultaneously lose to a global weight.
On KuaiRand-27K the attention head posts an identifiability gap of +128\% of the content gain yet trails the global weight by 0.7pp on PairAcc, and decoupling its gate input shows the inflation is architectural.
A positive shuffle gap alone can therefore look convincing while utility over a global weight is absent~\cite{chen2024m3csr,dong2025margo}, so we propose reporting both contrasts together as the audit standard.


\section{Related Work}

\subsection{Per-User Modality Weighting}
Per-user modality weighting has become a recurring design pattern in multimodal recommendation, motivated by the intuition that users differ in how much they rely on visual, textual, or acoustic content.
M3CSR learns a user modality-strength vector so that each user's collaborative profile scales the three content channels independently, and reports billion-user deployment as evidence of practical value~\cite{chen2024m3csr}.
MARGO conditions the per-user weight on modality-reliability estimates, arguing that weights optimized without reliability guidance may not reflect true user preferences~\cite{dong2025margo}.
The same intuition drives graph-based recommenders that build per-user modality-specific interaction graphs, routing message passing through whichever channel each user's history emphasizes~\cite{wei2019mmgcn,wang2023dualgnn}.
It also appears as user hypergraph structures that separate local modality preference from global item semantics~\cite{guo2024lgmrec}, as bilinear attention gates and meta-weight hypernetworks~\cite{shu2019metaweightnet} that condition the weight on the user's collaborative embedding.
Our audit focuses on the gate- and weight-vector subfamily, where the per-user weight is an explicit scalar or vector applied to modality channels.
Graph-based routing methods share the same intuition but differ structurally and fall outside the scope of this audit.
Across this family, the shared evaluation practice is to report a ranking gain over a no-modality or globally weighted baseline and attribute it to user-specific modality preference.
To our knowledge these evaluations provide no control that breaks the user-weight binding, so a gain attributable to a single global weight or to extra model capacity is indistinguishable from a genuinely user-specific signal.
Our audit supplies that control and shows it is necessary but not sufficient, since a shuffle control can certify a head as strongly user-specific while it simultaneously loses to a global weight.

\subsection{Multimodal Recommendation}
The broader multimodal recommendation literature fuses content features into collaborative backbones, and the fusion gain itself is well established.
Collaborative filtering backbones range from matrix factorization~\cite{koren2009mf} to graph-based propagation models such as NGCF~\cite{wang2019ngcf} and LightGCN~\cite{he2020lightgcn}; our audit uses this family as the shared backbone so that only the weighting head varies.
VBPR~\cite{he2016vbpr} extends BPR with visual item embeddings and demonstrates that content features improve ranking over interaction-only models.
Subsequent work mines latent item-item structure from multimodal features~\cite{zhang2021lattice,zhang2023micro} and propagates it with graph attention or convolution~\cite{tao2020mgat,wei2020grcn,zhou2023freedom,zhou2023dragon}, bootstraps contrastive views across modalities without auxiliary graphs~\cite{zhou2023bm3}, adds self-supervised and multi-view objectives over content and structure~\cite{tao2023slmrec,wei2023mmssl,yu2023mgcn,xu2025mentor}, and fuses spectrum-based modality representations to suppress cross-modality noise~\cite{ong2025smore}.
Recent surveys catalogue the rapid growth of this design space~\cite{zhou2023mmrecsurvey}.
This fusion gain is uneven across ranking metrics, because raw modality features carry noise that dedicated denoising and graph-refinement methods are designed to suppress~\cite{wei2020grcn,zhou2023freedom}.
Our own results later exhibit the same tension, where a single global weight improves pairwise accuracy and recall yet trades off NDCG on some corpora.
These methods improve recommendation by better representing items; none claims the modality weight must be per-user, the narrower claim our audit targets rather than the value of multimodal content in general.

\subsection{Permutation and Control Tests}
Permutation and control tests are a standard instrument for isolating entity-specific signal in machine learning.
The idea originates in permutation feature importance for random forests, where a feature is shuffled and the resulting loss increase measures its contribution~\cite{breiman2001rf}, and permutation likewise underpins distribution-free significance tests for whether a measured performance exceeds chance~\cite{ojala2010permutation}.
Fisher, Rudin, and Dominici~\cite{fisher2019mcr} generalize this into model reliance, the expected loss increase when a feature is permuted across an entire class of predictors, a model-agnostic measure that needs no retraining.
A parallel line in interpretability shows why such controls are indispensable, since sanity checks reveal many saliency methods are insensitive to the signal they claim to explain~\cite{adebayo2018sanity}, and controlled ablation benchmarks separate genuine importance from estimator artifacts~\cite{hooker2019benchmark}.
Our identifiability gap, real-shuf, follows this tradition by freezing the trained head and permuting the user-weight binding at evaluation within activity deciles, so the gap measures how much the binding matters net of activity-level confounds.
The key finding of our audit is that this control, while necessary, is not sufficient, since a gate reading the shared collaborative embedding inflates the shuffle gap through an architectural confound.
This mirrors a known hazard in variable-importance estimation, where permuting an input correlated with others biases importance upward unless the permutation is made conditional~\cite{strobl2008conditional}, and unrestricted permutation forces the model to extrapolate off the data manifold, so an honest estimate needs at least one additional model~\cite{hooker2021permutation}.
Decoupling the gate input causally confirms this inflation, which to our knowledge has not been identified or controlled for in prior work on per-user modality weighting.

\subsection{Audit Studies and Negative Results}
Our work joins a line of audit studies that re-examine published gains under matched baselines and controls.
Ferrari Dacrema et al.\ found that 11 of 12 reproducible neural recommendation methods were outperformed by properly tuned nearest-neighbor or matrix-factorization baselines~\cite{ferraridacrema2019progress}, and later showed the problem persists across conference years~\cite{ferraridacrema2021troubling}.
Rendle et al.\ demonstrated that a carefully tuned dot product matches or beats learned similarity functions once baselines are given equal care~\cite{rendle2020ncfvsmf}, echoing earlier evidence that the evaluation protocol alone can reorder top-N rankings~\cite{cremonesi2010topn}.
Reproducibility-driven analyses reach the same verdict for graph collaborative filtering, where reported gains shrink under fair comparison~\cite{anelli2023myth}, and for metric learning, where a reality check erases most claimed progress~\cite{musgrave2020reality}.
These studies, alongside broader commentary on how weak baselines and conflated gains distort measured progress~\cite{lipton2019troubling,sculley2018winners}, ask whether reported improvements survive fair comparison.
We ask a more specific question, namely whether the personalization mechanism in per-user modality weighting is real, and answer it with a paired-contrast criterion applicable to any weighting head on a shared backbone.
Unlike those audits, which largely rely on single runs without significance testing, our conclusions rest on five random seeds, paired $t$-tests, and Benjamini--Hochberg FDR correction across all six heads.
We further distinguish our audit targets from methods that improve multimodal recommendation through graph denoising~\cite{qi2025even} or calibration distillation~\cite{li2025guider}, since those methods address interaction noise rather than per-user modality preference, and their gains do not depend on the user-weight binding our audit tests.


\section{Audit Design}

\subsection{Datasets and Modality Features}
We audit on three short-video corpora with implicit watch-ratio feedback that differ in scale, modality structure, and platform origin.

\textbf{Tsinghua ShortVideo (TSV)}~\cite{shang2025shortvideo} is a short-video platform log filtered to users with at least 30 exposures, yielding 4,099 active users across three modality channels: a 256-dimensional visual embedding from mean-pooled ResNet/ViT frame features (L2-normalized), a 128-dimensional text embedding from ASR transcripts and titles reduced by truncated SVD, and a metadata channel encoding item category and author popularity.

\textbf{KuaiRand-27K}~\cite{gao2022kuairand} is a randomly exposed short-video log from 27,285 users; after compacting the item space to the 2.9M unique videos seen in training, we extract three modality channels from 58-dimensional tag features, 39-dimensional music features, and 22-dimensional metadata features.

\textbf{MicroLens-100K}~\cite{ni2023microlens} is a short-video corpus from a distinct UGC platform with 100,000 users and 19,738 items; modality channels are a 1,024-dimensional CLIP-RN50 visual embedding and a 1,024-dimensional BGE-M3 text embedding, both L2-normalized.

For all three corpora we split each user's interaction history by chronological order, assigning the first third to training, the middle third to validation, and the final third to test.
Watch ratio is the implicit preference proxy throughout; duration is excluded from all modality channels to prevent mechanical coupling with the label.
A fourth corpus, the Amazon-Baby e-commerce log, replicates the full audit outside the short-video domain and is reported in the Appendix.

\subsection{One Backbone, Six Weighting Heads}
We reduce the entire family onto one matrix-factorization collaborative backbone~\cite{koren2009mf} so that six weighting heads differ only in how the per-user modality weight is produced.
The backbone scores a user-item pair as $\hat{y}_{ui}=\mathbf{p}_u^\top\mathbf{q}_i+b_i+\sum_{m} w_{u,m}\,s_m(u,i)$, where $s_m(u,i)$ is the cosine similarity between a modality-specific item embedding and a shared item embedding, scaled by a learnable factor to keep content from overwhelming the collaborative signal~\cite{he2016vbpr}.
Two baselines bound the family: BPR~\cite{rendle2009bpr} sets all $w_{u,m}$ to zero, and GM learns a single global weight $w_m$ shared across users.
GM is not an unpersonalized baseline, since it keeps every user's own collaborative embedding $\mathbf{p}_u$ and personalizes what each user likes; the one quantity it does not make per-user is the cross-modal weight $w_m$, the variable this audit isolates.
Real-GM therefore asks whether per-user modality weighting adds anything on top of the collaborative personalization GM already provides.
Every per-user head is initialized to the uniform softmax so that all heads equal GM at initialization and any gap is purely learned.
Unlike Ferrari Dacrema et al.~\cite{ferraridacrema2019progress}, we audit our own reimplementations rather than original author code.
This is deliberate, since one shared backbone controls for dataset, training procedure, and evaluation protocol at once, so any measured gap is attributable to the weight-production mechanism rather than to backbone capacity or training setup.
Each head follows its published description as closely as the shared backbone permits, and the within-backbone comparisons stay internally valid because every contrast is paired on the same backbone, seeds, and metric.

The six heads differ in how $w_{u,\cdot}$ is produced, as summarized in Table~\ref{tab:heads}.
PUM learns a free per-user weight table with $U\times M$ parameters.
ATT applies a bilinear gate to the shared collaborative embedding: $w_{u,\cdot}=\text{softmax}(\mathbf{p}_u^\top W_\text{att})$.
MWN feeds the shared embedding into a two-layer hypernetwork: $w_{u,\cdot}=\text{softmax}(\text{MLP}(\mathbf{p}_u))$.
LRG uses a rank-8 independent per-user embedding $\mathbf{g}_u\in\mathbb{R}^8$ combined with $M$ shared basis vectors to produce weights, so the gate reads no collaborative signal.
The decoupled variants ATT$_d$ and MWN$_d$ replace $\mathbf{p}_u$ in the gate with a private embedding $\tilde{\mathbf{p}}_u$ trained separately and never shared with the collaborative path.
This lets us test whether the gate input, rather than the weighting itself, drives any observed effects.

\begin{table}[t]
\centering
{\small
\begin{tabular}{llc}
\toprule
Head & Weight mechanism & Reads $\mathbf{p}_u$? \\
\midrule
PUM   & Free $U\!\times\!M$ table, row softmax          & No  \\
ATT   & Bilinear: $\text{softmax}(\mathbf{p}_u^\top W_\text{att})$ & Yes \\
MWN   & Hypernetwork: $\text{softmax}(\text{MLP}(\mathbf{p}_u))$   & Yes \\
LRG   & Low-rank: $\text{softmax}(\mathbf{g}_u^\top G)$, rank 8    & No  \\
ATT$_d$ & Same as ATT; gate reads $\tilde{\mathbf{p}}_u$ & No  \\
MWN$_d$ & Same as MWN; MLP reads $\tilde{\mathbf{p}}_u$  & No  \\
\bottomrule
\end{tabular}}
\caption{Six per-user weighting heads audited on the shared backbone. The rightmost column marks whether the gate reads the shared collaborative embedding $\mathbf{p}_u$, which is the key variable in the decoupling analysis.}
\label{tab:heads}
\end{table}

\subsection{Two Paired Contrasts}
Each head faces two paired contrasts: the \emph{utility gap} real-GM asks whether a per-user weight helps at all, and the \emph{identifiability gap} real-shuf asks whether the weight is bound to the right user.
Writing $\mathrm{Perf}(\cdot)$ for a ranking metric, the utility gap is $\Delta_U(h)=\mathrm{Perf}(h_\text{real})-\mathrm{Perf}(\text{GM})$ and the identifiability gap is $\Delta_I(h)=\mathrm{Perf}(h_\text{real})-\mathrm{Perf}(h_\text{shuf})$.
A positive $\Delta_I$ certifies that the user-weight binding matters, yet it does not imply a positive $\Delta_U$, since only $\Delta_U>0$ shows the binding adds utility over the strongest non-personalized alternative.
The utility gap compares a trained head against GM, the strongest single-weight baseline, so a positive gap means personalization beats a global weight rather than merely beating no modality at all.
The identifiability gap compares the trained head against a version where each user's weight is replaced by that of a different user drawn from the same activity decile, averaged over five draws.

We constrain the shuffle to the same activity decile because activity level shapes the collaborative embedding independently of modality preference, and cross-decile swaps would confound activity-driven differences with the user-modality binding under test.
The shuffle must also happen after training, because a permutation applied during training is a bijection that the model absorbs through its user embeddings, leaving real and shuffled runs indistinguishable.
We therefore freeze each trained head and permute the weights only at evaluation, so the identifiability gap measures how much the learned user-weight binding matters, net of all activity effects~\cite{fisher2019mcr}.

\subsection{Instrument Calibration by Signal Implants}
To confirm the harness can detect user-specific structure when it exists, we implant synthetic modality preferences at four strengths and require a monotone dose-response.
For each user we draw a latent preferred modality uniformly at random, then amplify the content score of that modality by a factor $\alpha\in\{0.25,0.5,0.75,1.0\}$ during training and evaluation.
We track capture AUROC, the area under the ROC curve for predicting the planted preferred modality from the learned weight vector, alongside the resulting identifiability gap, and require both to rise monotonically with $\alpha$ (Figure~\ref{fig:decouple}b).

This calibration also yields an effect-size ceiling, since $\alpha=0.25$, the weakest implant we can already detect, yields capture AUROC 0.57 on Tsinghua ShortVideo and 0.64 on KuaiRand-27K.
Any natural signal weaker than this cannot explain the null utility result, so the null is a genuine absence of user-specific signal at this scale rather than a measurement failure.
The same monotone dose-response holds on all four corpora (Appendix).
The calibration uses a single seed, since the relevant property is the monotone ordering across four implant strengths.

\subsection{Statistical Protocol}
All conclusions rest on five random seeds per configuration and paired $t$-tests on per-seed aggregates, a statistical standard not previously applied to this family of claims~\cite{ferraridacrema2019progress}.
We evaluate three metrics: pairwise ranking accuracy on held-out exposure pairs (PairAcc), which most directly probes modality-driven preference because pairs ranked differently by the two modalities are where a user-specific weight should help most; together with NDCG@20 and Recall@20 for ranking quality.
We express each gap in percentage points and, in the figures, as a fraction of the PairAcc content gain, so that identifiability and utility share one interpretable scale.
We apply Benjamini--Hochberg FDR correction across the six heads for each metric and dataset; main tables report paired-$t$ significance for readability, and conclusions are robust to the correction.
The identifiability gaps are stable under twenty eval-time permutations, changing by at most 0.11pp against the five-draw setting, so five draws report a converged gap (Appendix).
A head clears the personalization bar only when it beats GM on real-GM, not merely when it beats its own shuffle on real-shuf.


\section{Results}

\subsection{A Single Global Weight Captures the Content Gain}
Across all three corpora, a single global modality weight already delivers nearly all of the content gain: +1.9pp on Tsinghua ShortVideo, +3.6pp on KuaiRand-27K, and +3.5pp on MicroLens-100K over the no-modality baseline (all $p<.001$).
We measure this gain on pairwise ranking accuracy (PairAcc) over held-out exposure pairs, the metric that most directly probes modality-driven preference because it scores the pairs the two modalities can rank differently.
The same global weight also raises Recall on all three corpora and trades off NDCG on some, reflecting a modality-collaborative tension that varies across corpora (Table~\ref{tab:contentgain}).
This tension cannot bias the audit, since every contrast is paired and affects both arms equally.
The content gain is real, and the audit asks only whether making the weight per-user adds anything beyond it.
Throughout, we report each per-user gap as a fraction of this PairAcc content gain.

\begin{table}[t]
\centering
{\small
\begin{tabular}{lccc}
\toprule
Dataset & PairAcc & NDCG & Recall \\
\midrule
Tsinghua ShortVideo & $+$1.89$^{***}$ & $-$0.40 & $+$1.21$^{*}$ \\
KuaiRand-27K & $+$3.61$^{***}$ & $-$2.46$^{***}$ & $+$2.64$^{***}$ \\
MicroLens-100K & $+$3.50$^{***}$ & $+$1.83$^{***}$ & $+$5.86$^{***}$ \\
\bottomrule
\end{tabular}}
\caption{Content gain of a single global modality weight over the no-modality baseline (GM$-$BPR, percentage points). The gain concentrates on pairwise accuracy and recall; a global weight trades off NDCG on some corpora, a known modality-collaborative tension. Five seeds, paired $t$-tests. $^{*}p<.05$, $^{**}p<.01$, $^{***}p<.001$.}
\label{tab:contentgain}
\end{table}

\subsection{Per-User Weighting Adds No Consistent Utility}
Making the weight per-user adds no consistent utility in our evaluation. No implementation wins on all three corpora across all three metrics, and the few positive gaps are small ($\leq$0.9pp) and flip across corpora and metrics.
On Tsinghua ShortVideo every head loses to the global weight on NDCG and Recall, and five of six lose on PairAcc (Table~\ref{tab:utility}).
On KuaiRand-27K the meta-weight family is the lone winner, small but significant, up to +0.9pp on NDCG, yet the same heads are negative across all three metrics on Tsinghua ShortVideo and MicroLens-100K.
On MicroLens-100K all six heads are negative across all three metrics with no exception.
The decoupling analysis below shows the few surviving positive cells are indifferent to the user-weight binding.
GM is not a weakened baseline. A capacity-matched global gate that spends the meta-weight budget on item content rather than user identity (GMc) does not help; it loses 2.84pp of PairAcc on Tsinghua ShortVideo and 3.75pp on MicroLens-100K. On KuaiRand-27K it stays flat on PairAcc while its ranking metrics collapse by 7.59pp NDCG and 16.77pp Recall, so extra capacity does not rescue the global weight (Appendix).

\begin{table*}[t]
\centering
{\small
\begin{tabular}{lcccccccccc}
\toprule
& \multicolumn{3}{c}{Tsinghua ShortVideo} & \multicolumn{3}{c}{KuaiRand-27K} & \multicolumn{3}{c}{MicroLens-100K} \\
\cmidrule(lr){2-4}\cmidrule(lr){5-7}\cmidrule(lr){8-10}
Head & PairAcc & NDCG & Recall & PairAcc & NDCG & Recall & PairAcc & NDCG & Recall \\
\midrule
PUM & $-$0.21 & $-$0.12$^{**}$ & $-$0.20 & $-$0.44$^{***}$ & $-$0.32$^{***}$ & $-$0.60$^{***}$ & $-$0.13$^{*}$ & $-$0.24$^{*}$ & $-$0.31$^{*}$ \\
ATT & $+$0.44 & $-$0.46$^{***}$ & $-$0.53$^{**}$ & $-$0.69$^{**}$ & $+$0.54$^{***}$ & $-$0.17 & $-$1.25$^{***}$ & $-$0.97$^{***}$ & $-$1.83$^{***}$ \\
MWN & $-$0.17 & $-$0.63$^{**}$ & $-$1.08$^{**}$ & $+$0.29 & $+$0.88$^{***}$ & $+$0.57$^{**}$ & $-$0.91$^{***}$ & $-$1.00$^{***}$ & $-$1.39$^{***}$ \\
LRG & $-$0.42 & $-$0.56$^{**}$ & $-$0.85$^{**}$ & $-$0.48$^{***}$ & $-$0.35$^{***}$ & $-$0.60$^{***}$ & $-$1.12$^{***}$ & $-$0.88$^{***}$ & $-$1.74$^{***}$ \\
ATT$_d$ & $-$0.70 & $-$0.65$^{**}$ & $-$1.14$^{**}$ & $-$0.50$^{***}$ & $-$0.36$^{***}$ & $-$0.62$^{***}$ & $-$1.55$^{***}$ & $-$1.17$^{***}$ & $-$2.40$^{***}$ \\
MWN$_d$ & $-$0.69 & $-$0.68$^{**}$ & $-$1.31$^{**}$ & $+$0.26$^{**}$ & $+$0.16$^{**}$ & $+$0.32$^{***}$ & $-$0.89$^{**}$ & $-$0.80$^{*}$ & $-$1.26$^{**}$ \\
\bottomrule
\end{tabular}}
\caption{Utility gap of per-user modality weighting against a single global weight (real-GM, percentage points) across six heads, three corpora, and three metrics. No head wins on all three corpora across all three metrics; positive cells are small and flip. Subscript $d$ marks decoupled-gate variants (Audit Design). Five seeds, paired $t$-tests; stars denote Benjamini--Hochberg FDR $q$-values across the six heads within each column. $^{*}q<.05$, $^{**}q<.01$, $^{***}q<.001$.}
\label{tab:utility}
\end{table*}

\subsection{Identifiability Dissociates from Utility}
\begin{figure}[t]
\centering
\includegraphics[width=0.9\columnwidth]{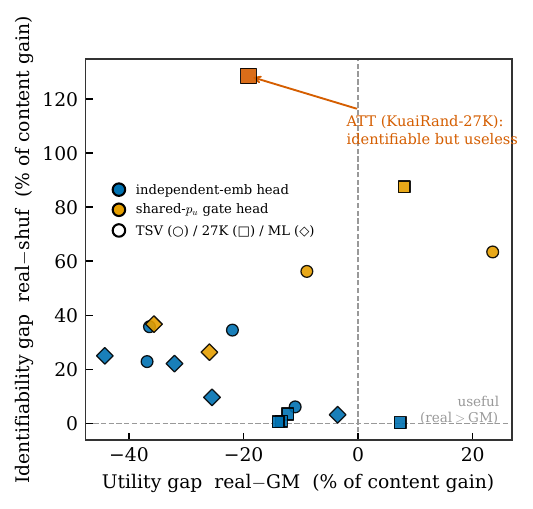}
\caption{Utility versus identifiability for six heads on three corpora (circle=TSV, square=27K, diamond=MicroLens), each gap as a fraction of the PairAcc content gain. Points in the top-left are identifiable (large real-shuf) yet useless (negative real-GM); the KuaiRand-27K attention head is the extreme case. Large gaps track shared-$p_u$ gate heads, not utility.}
\label{fig:dissociation}
\end{figure}

Identifiability dissociates sharply from utility. The attention head on KuaiRand-27K gains +4.6pp from its own user-weight binding, +128\% of the content gain, while losing to the global weight by 0.7pp.
Table~\ref{tab:identifiability} reports the identifiability gaps, and Figure~\ref{fig:dissociation} plots every head across all three corpora in the (utility, identifiability) plane, where this head sits alone in the top-left, strongly identifiable yet below the global weight.
The large identifiability gaps concentrate in heads whose gate reads the shared collaborative embedding (ATT, MWN), not the independent-embedding heads (PUM, LRG).
A shuffle control alone would certify the top-left heads as strongly user-specific, yet real-GM shows they carry no utility, so the two contrasts measure different things and only real-GM tracks whether personalization helps.

\begin{table}[t]
\centering
{\small
\setlength{\tabcolsep}{3pt}
\begin{tabular}{lcccccc}
\toprule
& \multicolumn{2}{c}{TSV} & \multicolumn{2}{c}{KuaiRand-27K} & \multicolumn{2}{c}{MicroLens-100K} \\
\cmidrule(lr){2-3}\cmidrule(lr){4-5}\cmidrule(lr){6-7}
Head & pp & \%gain & pp & \%gain & pp & \%gain \\
\midrule
PUM & $+$0.12 & 6.1 & $+$0.13$^{***}$ & 3.5 & $+$0.11$^{***}$ & 3.2 \\
ATT & $+$1.20$^{*}$ & 63.4 & $+$4.64$^{***}$ & 128.5 & $+$1.29$^{***}$ & 36.7 \\
MWN & $+$1.06$^{**}$ & 56.2 & $+$3.15$^{**}$ & 87.5 & $+$0.92$^{***}$ & 26.4 \\
LRG & $+$0.65$^{*}$ & 34.5 & $+$0.03$^{**}$ & 0.9 & $+$0.78$^{***}$ & 22.1 \\
ATT$_d$ & $+$0.43 & 22.9 & $+$0.02$^{*}$ & 0.6 & $+$0.88$^{***}$ & 25.0 \\
MWN$_d$ & $+$0.68$^{**}$ & 35.8 & $+$0.01$^{*}$ & 0.2 & $+$0.34$^{**}$ & 9.6 \\
\bottomrule
\end{tabular}}
\caption{Identifiability gap (real-shuf on PairAcc): percentage points and share of the content gain across three corpora (TSV is Tsinghua ShortVideo). Large gaps track the shared-$p_u$ gate heads (ATT, MWN), not the heads closest to utility. Five seeds, five eval-time permutations, paired $t$-tests; stars denote Benjamini--Hochberg FDR $q$-values across the six heads within each corpus. $^{*}q<.05$, $^{**}q<.01$, $^{***}q<.001$.}
\label{tab:identifiability}
\end{table}

\subsection{Decoupling and Dose-Response Close the Loop}
\begin{figure}[t]
\centering
\includegraphics[width=0.86\columnwidth]{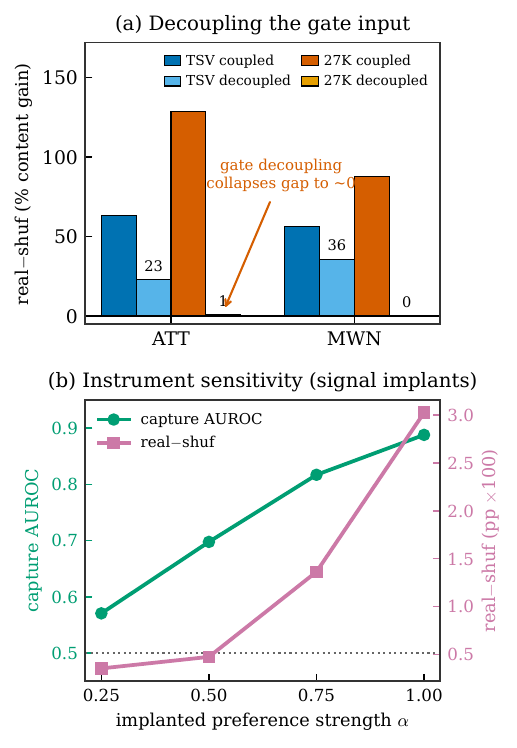}
\caption{(a) Decoupling the gate input from the shared collaborative embedding collapses the identifiability gap, most sharply on KuaiRand-27K where it falls to near zero; the utility conclusion is unchanged. (b) Signal-implant calibration: capture AUROC and the identifiability gap both rise monotonically with implant strength $\alpha$, confirming the harness detects user-specific structure when it exists; the same monotone pattern holds on all four corpora (Appendix).}
\label{fig:decouple}
\end{figure}

Decoupling the gate input from the collaborative embedding collapses the inflated identifiability gaps to near zero on KuaiRand-27K, from +128\% to +0.6\% for attention and from +87\% to +0.2\% for meta-weight, while the utility conclusion stands.
On Tsinghua ShortVideo the same intervention shrinks both gates far less, and the decoupled gaps stay clearly nonzero (Figure~\ref{fig:decouple}a).
For the single positive PairAcc cell, Tsinghua attention, decoupling flips the gap negative (from +0.44 to $-$0.70pp, $p=.038$), removing that exception to the null.
The KuaiRand-27K meta-weight cells shrink under decoupling yet stay positive, but a gain that survives swapping users' weights is capacity, not personalization.
Figure~\ref{fig:decouple}b calibrates the harness, where implanting synthetic per-user preferences at four strengths raises capture AUROC monotonically from 0.57 to 0.89 on Tsinghua ShortVideo and from 0.64 to 1.00 on KuaiRand-27K.
The identifiability gap rises in parallel from +0.35pp to +3.0pp, so the null utility result reflects an absent signal, not a blind instrument, and the same monotone dose-response holds on all four corpora (Appendix).
All four results replicate on Amazon-Baby, a cross-domain e-commerce corpus, where every utility gap is negative and no identifiability gap is inflated (Appendix).
The dissociation also reproduces under a LightGCN backbone rather than matrix factorization, where per-user heads still fail to beat the global weight while their weights stay identifiable, so our audit spans two collaborative backbones (Appendix).

\section{Why Per-User Weighting Shows No Consistent Utility, and What Would Work}

The measurable value of a per-user weight is, in our audited settings, empirically bounded by a global weight plus model capacity, and two mechanisms explain why.

The first reason is informational, since a scalar watch ratio never records whether the frames or the caption earned it, so implicit feedback, which encodes only a confidence-weighted preference per interaction~\cite{hu2008implicit}, cannot assign credit across modalities.
The utility gap therefore sits at the global weight regardless of head capacity.
The dose-response calibration confirms a genuine ceiling, not a measurement failure, since capture AUROC rises monotonically with implant strength while natural data stays at the null.
A cold-start slice sharpens the account: the coldest activity decile, where any surviving per-user signal should surface, recovers no benefit on either corpus and is significantly negative on KuaiRand-27K, so the most plausible niche for per-user weighting is empty (Appendix).

The second mechanism explains the inflated identifiability gaps rather than the utility gaps: a gate that reads the shared collaborative embedding routes user identity through a high-capacity path, so replacing one user's weight with another's disturbs that path and makes the shuffle look damaging even when the weight carries no modality signal.
Decoupling the gate input collapses the inflated real-shuf to near zero while the utility conclusion is unchanged, confirming the large identifiability gap measures absorbed capacity, not a user-specific modality signal.

These two mechanisms yield three prescriptions.
A shuffle-only standard would be self-reinforcing, since a positive real-shuf is easy to obtain and sufficient for publication, so each paper reporting it alone would license the next, the same propagation mechanism Ferrari Dacrema et al.\ identified for weak baselines~\cite{ferraridacrema2019progress}.
For auditors, a positive real-shuf without a positive real-GM is not evidence of useful personalization, so we argue that a claim of effective per-user modality weighting should report real-GM $>0$ with statistical significance on at least two independent corpora, as a minimum evidentiary standard, and should calibrate the harness with a signal implant before trusting any null.
For method designers, a genuine per-user modality weight may require supervision that scalar feedback does not provide, such as missing-modality natural experiments, counterfactual exposure logs, or session-level interaction traces.
For parameterization, a weight could condition on interpretable user-level modality-reliability statistics rather than a free identity embedding.

More broadly, our audit operationalizes a distinction between identity-sensitive parameterization and useful personalization reaching past multimodal recommendation: any architecture conditioning parameters on a user embedding can inflate an identifiability signal through capacity alone, so separating a binding that merely matters from one that adds utility is a general requirement for personalization claims.

\section{Limitations}
Our audit is bounded in three ways.
It covers weighting-class personalization with a static user-weight binding, not fusion architectures or session-level dynamic weights, and the reduction to one backbone assumes the family shares its behavior in the head rather than the collaborative path.
A LightGCN backbone check supports this assumption (Results; Appendix).
By design we audit reimplementations on one shared backbone, not original author code, so our claims concern the weight-production mechanism under matched conditions rather than any original system's reported numbers.
Its main results rest on three short-video corpora with implicit-feedback proxy metrics, so pairwise accuracy stands in for a modality-driven preference we cannot observe directly; the e-commerce replication widens the domain but shares the proxy.
The cold-start slice ruling out a low-activity niche is powered mainly by KuaiRand-27K, since the Tsinghua ShortVideo deciles are too small to settle that question alone.

\section{Conclusion}
Across six implementations, three independent short-video corpora, and a cross-domain e-commerce replication, the measurable value of per-user modality weighting is, in these settings, empirically bounded by a global modality weight plus model capacity, not a user-specific signal.
A personalization claim is credible only when a head beats the global weight, not merely its own shuffle, so both contrasts must be reported together as a minimum evidentiary standard.
We release the audit harness, the two-contrast protocol, and all per-seed runs as supplementary material.

\section{Ethical Statement}
This work is a methodological audit of existing recommendation techniques and introduces no new deployed system.
It uses three short-video logs and one e-commerce review log under their intended research terms, reporting only aggregate ranking statistics with no personally identifying content or human-subject intervention.
By showing that a common personalization claim does not survive a controlled contrast, it raises the evidentiary bar rather than guiding user targeting or profiling.


\appendix
\section{Technical Appendix}

This appendix gives the configuration needed to reproduce the audit, together with robustness analyses: a capacity-matched global baseline, shuffle-permutation stability, cold-start slices, a cross-domain replication on Amazon-Baby, a LightGCN backbone check, and dose-response calibration on all four corpora.
We release the full harness and every per-seed run; the tables below list the static settings shared across all runs.

\subsection{Backbone and Weighting Heads}
All heads share one matrix-factorization backbone with user and item embeddings of dimension $d=64$ and a per-item bias.
The modality term is a bounded gate: each modality contributes a cosine similarity scaled into $[-c,c]$ by a single global learnable factor $c$, so content corrects the collaborative score instead of overwhelming it, and this scale is never per-user.
Every per-user head zeroes its output layer at initialization, so all heads equal GM at step zero and any gap is purely learned.
Table~\ref{tab:apphead} lists the head-specific parameterizations; PUM, LRG, ATT$_d$, and MWN$_d$ never read the shared collaborative embedding $\mathbf{p}_u$, while ATT and MWN do.

\begin{table}[t]
\centering
{\small
\begin{tabular}{lll}
\toprule
Head & Parameters & Gate input \\
\midrule
BPR    & none ($w_{u,m}{=}0$)                       & n/a \\
GM     & one global $w\in\mathbb{R}^{M}$            & n/a \\
PUM    & free table $U\!\times\!M$, zero init        & n/a \\
ATT    & $W_q\in\mathbb{R}^{d\times d}$, $K\in\mathbb{R}^{M\times d}$ & $\mathbf{p}_u$ \\
MWN    & MLP $64\!\to\!64\!\to\!M$, ReLU             & $\mathbf{p}_u$ \\
LRG    & $\mathbf{g}_u\in\mathbb{R}^{8}$, basis $\mathbb{R}^{8\times M}$ & $\mathbf{g}_u$ \\
ATT$_d$ & as ATT, private gate embedding $\tilde{\mathbf{p}}_u$ & $\tilde{\mathbf{p}}_u$ \\
MWN$_d$ & as MWN, private gate embedding $\tilde{\mathbf{p}}_u$ & $\tilde{\mathbf{p}}_u$ \\
\bottomrule
\end{tabular}}
\caption{Head parameterizations on the shared $d=64$ backbone. $U$ users, $M$ modalities ($M{=}3$ on Tsinghua ShortVideo and KuaiRand-27K, $M{=}2$ on MicroLens-100K and Amazon-Baby). Output layers are zero-initialized so every head starts equal to GM.}
\label{tab:apphead}
\end{table}

\subsection{Optimization and Final Hyperparameters}
We use the same configuration for every head on every corpus, with no per-head tuning, so that any measured gap reflects the weighting mechanism rather than a search budget.
Table~\ref{tab:apphp} lists the final values.
Weights are optimized with Adam under a BPR pairwise objective.

\begin{table}[t]
\centering
{\small
\begin{tabular}{ll}
\toprule
Setting & Value \\
\midrule
Embedding dimension $d$        & 64 \\
LRG rank                       & 8 \\
Optimizer                      & Adam \\
Learning rate                  & $1\times10^{-2}$ \\
Weight decay                   & $1\times10^{-6}$ \\
Batch size                     & 8192 \\
Training epochs                & 15 \\
Embedding init                 & $\mathcal{N}(0,0.01^2)$ \\
Random seeds                   & $\{0,1,2,3,4\}$ \\
Shuffle draws $n_{\text{perm}}$ & 5 (within-decile) \\
Activity deciles               & 10 \\
Ranking cutoff $K$             & 20 \\
Sampled negatives (eval)       & 99 \\
Implant strengths $\alpha$     & $\{0.25,0.5,0.75,1.0\}$ \\
\bottomrule
\end{tabular}}
\caption{Final hyperparameters, shared across all heads and all corpora.}
\label{tab:apphp}
\end{table}

\subsection{Dataset Cards}
The three short-video corpora use watch ratio as the implicit preference proxy; Amazon-Baby binarizes rating interactions into implicit feedback.
Duration is excluded from every modality channel to prevent mechanical coupling with the label.
Each user's history is split into training, validation, and test by chronological order, taking the first, middle, and final thirds.

\textbf{Tsinghua ShortVideo (TSV)}~\cite{shang2025shortvideo}. A short-video platform log filtered to users with at least 30 exposures, yielding 4,099 active users.
Three channels: a 256-dimensional L2-normalized visual embedding from mean-pooled ResNet/ViT frame features, a 128-dimensional text embedding from ASR transcripts and titles reduced by truncated SVD, and a metadata channel encoding item category and author popularity.

\textbf{KuaiRand-27K}~\cite{gao2022kuairand}. A randomly exposed short-video log from 27,285 users.
We compact the item space to the 2.9M unique videos seen in training and extract 58-dimensional tag, 39-dimensional music, and 22-dimensional metadata features.

\textbf{MicroLens-100K}~\cite{ni2023microlens}. A short-video corpus from a distinct UGC platform with 100,000 users and 19,738 items.
Two channels: a 1,024-dimensional CLIP-RN50 visual embedding and a 1,024-dimensional BGE-M3 text embedding, both L2-normalized.

\textbf{Amazon-Baby}~\cite{mcauley2015amazon}. An e-commerce review log with visual and textual product features, using the preprocessed release of FREEDOM~\cite{zhou2023freedom}: a 4,096-dimensional CNN image embedding and a 384-dimensional sentence-transformer text embedding, both L2-normalized.
Filtering to users with at least 10 interactions yields 4,382 users, 6,883 items, and 67,060 interactions over two modality channels.

\subsection{Evaluation and Statistical Protocol}
We report three metrics on the held-out test third: pairwise ranking accuracy on held-out exposure pairs (PairAcc), and NDCG@20 and Recall@20 against 99 sampled negatives per positive.
PairAcc most directly probes modality-driven preference because a user-specific weight should help most on pairs that the two modalities rank differently.
Because sampled ranking metrics can diverge from their full-ranking counterparts~\cite{krichene2020sampled}, we treat PairAcc over observed exposure pairs as the primary probe and read NDCG and Recall as corroborating evidence.
The identifiability gap real-shuf freezes each trained head and, only at evaluation, replaces each user's weight with that of another user drawn from the same activity decile, averaged over five draws.
Every reported number aggregates five seeds; significance uses paired $t$-tests on per-seed aggregates, with Benjamini--Hochberg FDR correction across the six heads for each metric and dataset.

\subsection{Capacity-Robustness of the Global Baseline}
A referee may ask whether GM is a weakened strawman that any per-user head could beat on capacity alone.
To test this, we add GMc, a global head that matches the meta-weight capacity but conditions the weight on item content rather than user identity: a two-layer MLP maps each item embedding to a softmax over modalities, shared across all users.
GMc adds exactly the parameters that MWN spends on personalization, but spends them on content, so a gap between them isolates capacity from identity.
Table~\ref{tab:gmc} reports GMc$-$GM on three corpora.
Extra capacity does not help the global weight; it hurts.
On Tsinghua ShortVideo GMc loses 2.84pp of PairAcc, and on KuaiRand-27K it collapses the ranking metrics by 7.59pp NDCG and 16.77pp Recall ($p<.001$), with PairAcc statistically flat ($-$0.23pp, $p=.30$).
On MicroLens-100K GMc loses 3.75pp of PairAcc, 2.92pp of NDCG, and 5.33pp of Recall (all $p<.001$).
The simple global weight is therefore near-optimal and robust to added capacity, not a strawman, which is why every per-user head is measured against GM rather than against a no-modality baseline.
We stress that the GMc collapse is not evidence that personalization is real: a content-conditioned gate that destabilizes ranking on three corpora is a poor model, not proof that a user-keyed gate is a good one.

\begin{table}[t]
\centering
{\small
\begin{tabular}{lc}
\toprule
Dataset & GMc$-$GM (PairAcc) \\
\midrule
Tsinghua ShortVideo & $-$2.84 \\
KuaiRand-27K & $-$0.23 \\
MicroLens-100K & $-$3.75 \\
\bottomrule
\end{tabular}}
\caption{Capacity-matched global baseline GMc against GM (GMc$-$GM, percentage points on PairAcc). GMc spends MWN-level capacity on item content rather than user identity. Added capacity does not help a global weight, confirming GM is near-optimal rather than a strawman. Five seeds, paired $t$-tests.}
\label{tab:gmc}
\end{table}

\subsection{Shuffle-Permutation Robustness}
The identifiability gaps in the main results average five eval-time permutations, so we verify that five draws suffice.
We rerun Tsinghua ShortVideo seeds 0 to 2 with twenty permutations and compare the gap and its per-permutation spread against the five-draw setting (Table~\ref{tab:nperm}).
The gap moves by at most 0.11pp across all six heads, and the per-permutation standard deviation is 0.17 to 0.38pp, far below the coupled-gate gaps near 1.2pp.
The large identifiability gaps are therefore a property of the user-weight binding, not permutation noise, and five draws already report a converged gap.

\begin{table}[t]
\centering
{\small
\begin{tabular}{lccc}
\toprule
Head & gap@20 (pp) & gap@5 (pp) & per-perm SD (pp) \\
\midrule
PUM     & $+$0.179 & $+$0.195 & 0.167 \\
ATT     & $+$1.258 & $+$1.150 & 0.383 \\
MWN     & $+$1.178 & $+$1.163 & 0.297 \\
LRG     & $+$0.721 & $+$0.706 & 0.349 \\
ATT$_d$ & $+$0.294 & $+$0.296 & 0.318 \\
MWN$_d$ & $+$0.675 & $+$0.617 & 0.340 \\
\bottomrule
\end{tabular}}
\caption{Identifiability gap under twenty versus five eval-time permutations (Tsinghua ShortVideo, seeds 0 to 2). Gaps change by at most 0.11pp; the per-permutation standard deviation is far below the coupled-gate gaps, so five draws suffice.}
\label{tab:nperm}
\end{table}

\subsection{Cold-Start Activity Slices}
If a per-user weight helped anywhere, it should help most for cold users whom the collaborative path cannot yet describe, so we slice the utility gap by user activity decile (d0 coldest, d9 hottest) at evaluation, without changing training.
Table~\ref{tab:coldstart} reports the coldest and hottest decile gaps and the paired significance at d0.
No head concentrates a positive utility gap at the cold end.
On Tsinghua ShortVideo the coldest decile is insignificant for all six heads ($p$ from .21 to .82), and the signs are scattered rather than monotone toward the cold end.
On KuaiRand-27K, where each decile holds roughly 2{,}700 users and power is ample, four of six heads are significantly negative at d0 ($p<.001$), so the coldest users fare worse, not better; the meta-weight heads are uniformly small-positive across all deciles with no cold concentration.
The most plausible niche for per-user weighting therefore does not exist on either corpus, which strengthens the informational account in the main text.

\begin{table}[t]
\centering
{\small
\begin{tabular}{lcccccc}
\toprule
& \multicolumn{3}{c}{Tsinghua ShortVideo} & \multicolumn{3}{c}{KuaiRand-27K} \\
\cmidrule(lr){2-4}\cmidrule(lr){5-7}
Head & d0 & d9 & $p_{d0}$ & d0 & d9 & $p_{d0}$ \\
\midrule
PUM     & $-$0.4 & $-$0.7 & .21 & $-$0.5 & $-$0.4 & $<$.001 \\
ATT     & $+$0.4 & $+$1.3 & .33 & $-$0.7 & $-$0.6 & $<$.001 \\
MWN     & $+$0.2 & $+$0.8 & .65 & $+$0.2 & $+$0.4 & .24 \\
LRG     & $-$0.5 & $-$0.1 & .25 & $-$0.5 & $-$0.4 & $<$.001 \\
ATT$_d$ & $-$0.1 & $-$0.2 & .82 & $-$0.5 & $-$0.4 & $<$.001 \\
MWN$_d$ & $+$0.8 & $-$0.2 & .44 & $+$0.2 & $+$0.3 & .002 \\
\bottomrule
\end{tabular}}
\caption{Utility gap (real-GM on PairAcc, percentage points) at the coldest (d0) and hottest (d9) activity deciles, with paired significance at d0. No head concentrates positive utility at the cold end; on KuaiRand-27K the coldest users are significantly worse. Five seeds.}
\label{tab:coldstart}
\end{table}

\subsection{Cross-Domain Replication: Amazon-Baby}
Every main-text conclusion replicates on Amazon-Baby, an e-commerce corpus outside the short-video domain, under the identical protocol: five seeds, six heads, both contrasts, and within-decile eval-time shuffles.
The global weight delivers the largest content gain of the four corpora, +6.97pp PairAcc, +5.61pp NDCG, and +9.26pp Recall (all $p<.001$), and it is the only corpus where all three metrics are positive.
Making the weight per-user helps nowhere: all 18 head-metric utility cells are negative, from $-$0.20 to $-$1.83pp, and 17 of 18 are significant under FDR (Table~\ref{tab:amazon}).
The identifiability gaps are all at most 4.1\% of the content gain, and the only gap passing FDR is ATT$_d$ at +0.15pp, negligible in size.
The coupled-gate inflation seen on the short-video corpora does not appear here, which is consistent with the architectural account: the confound strength scales predictably with how much identity information the collaborative embedding carries, and with 4,382 users and sparse histories the embedding carries little.
The cold-start slice is also flat: the coldest decile is insignificant for all six heads ($p$ from .17 to .95).
The corpus has two modality channels, image and text, and its dose-response calibration appears in the four-corpus table below.

\begin{table}[t]
\centering
{\small
\begin{tabular}{lccccc}
\toprule
& \multicolumn{3}{c}{Utility gap real-GM} & \multicolumn{2}{c}{real-shuf} \\
\cmidrule(lr){2-4}\cmidrule(lr){5-6}
Head & PairAcc & NDCG & Recall & pp & \%gain \\
\midrule
PUM     & $-$0.36$^{**}$ & $-$0.20$^{**}$ & $-$0.23$^{**}$ & $-$0.03 & $-$0.5 \\
ATT     & $-$0.76$^{**}$ & $-$0.54$^{**}$ & $-$0.97$^{***}$ & $+$0.28 & 4.1 \\
MWN     & $-$0.26 & $-$0.17$^{*}$ & $-$0.39$^{*}$ & $+$0.09 & 1.2 \\
LRG     & $-$0.94$^{***}$ & $-$0.66$^{***}$ & $-$1.20$^{***}$ & $+$0.08 & 1.2 \\
ATT$_d$ & $-$1.29$^{**}$ & $-$0.97$^{***}$ & $-$1.83$^{***}$ & $+$0.15$^{*}$ & 2.1 \\
MWN$_d$ & $-$1.05$^{**}$ & $-$0.78$^{***}$ & $-$1.62$^{**}$ & $-$0.03 & $-$0.4 \\
\bottomrule
\end{tabular}}
\caption{Amazon-Baby replication: utility gap (real-GM, percentage points) and identifiability gap (real-shuf, percentage points and share of the +6.97pp content gain). All utility cells are negative; no identifiability gap is inflated. Five seeds, paired $t$-tests; stars denote Benjamini--Hochberg FDR $q$-values across the six heads per column. $^{*}q<.05$, $^{**}q<.01$, $^{***}q<.001$.}
\label{tab:amazon}
\end{table}

\subsection{Backbone Robustness: LightGCN}
To test whether the audit conclusion depends on the matrix-factorization backbone, we replace it with a two-layer LightGCN: embeddings propagate over the symmetrically normalized user-item adjacency, the final embedding averages layers 0 to 2, and every head, contrast, and evaluation step is unchanged.
On Tsinghua ShortVideo with three seeds, the audit conclusion replicates: per-user heads still fail to beat the global weight, with ATT$-$GM at $-$0.41pp ($p=.018$) and MWN$-$GM at $-$0.49pp ($p=.079$), while their weights remain identifiable, with real-shuf at +0.76pp ($p=.054$) for ATT and +0.85pp ($p=.033$) for MWN (Table~\ref{tab:lgcn}).
The utility-identifiability dissociation is therefore not an artifact of the matrix-factorization backbone.
We scope this check to the audit conclusion: the content gain itself is metric-dependent under LightGCN, positive on Recall only, so we make no claim that the content gain replicates under this backbone.

\begin{table}[t]
\centering
{\small
\begin{tabular}{lcc}
\toprule
Contrast & Mean (pp) & $p$ \\
\midrule
ATT$-$GM        & $-$0.41 & .018 \\
MWN$-$GM        & $-$0.49 & .079 \\
ATT real$-$shuf & $+$0.76 & .054 \\
MWN real$-$shuf & $+$0.85 & .033 \\
\bottomrule
\end{tabular}}
\caption{Audit contrasts under a two-layer LightGCN backbone (Tsinghua ShortVideo, PairAcc, three seeds, one-sample $t$ on per-seed means). Per-user heads do not beat the global weight, yet their weights remain identifiable, replicating the dissociation.}
\label{tab:lgcn}
\end{table}

\subsection{Dose-Response Calibration on All Four Corpora}
The signal-implant calibration of the main text extends to all four corpora: capture AUROC rises monotonically with implant strength $\alpha$ on every corpus (Table~\ref{tab:dose}).
Each run implants a synthetic per-user preferred modality into the PUM head on a single seed, matching the main-text protocol; the relevant property is the monotone ordering, a qualitative instrument check.
On MicroLens-100K the $\alpha=0.25$ implant is at chance level, so the natural-gap anchor on that corpus is $\alpha=0.5$: natural identifiability gaps sit below what an $\alpha=0.5$ implant produces.
On Tsinghua ShortVideo and KuaiRand-27K the anchor is $\alpha=0.25$, as reported in the main text.

\begin{table}[t]
\centering
{\small
\begin{tabular}{lcccc}
\toprule
Corpus & $\alpha{=}0.25$ & $\alpha{=}0.5$ & $\alpha{=}0.75$ & $\alpha{=}1.0$ \\
\midrule
Tsinghua ShortVideo & .571 & .698 & .817 & .888 \\
KuaiRand-27K        & .638 & .898 & .990 & 1.000 \\
MicroLens-100K      & .501 & .537 & .655 & .896 \\
Amazon-Baby         & .511 & .592 & .789 & .974 \\
\bottomrule
\end{tabular}}
\caption{Capture AUROC of the signal implant at four strengths on all four corpora. AUROC rises monotonically with $\alpha$ everywhere, so the audit instrument detects user-specific structure whenever it is planted, and the natural nulls are not measurement failures. Single seed, PUM head.}
\label{tab:dose}
\end{table}

\subsection{Computing Infrastructure}
Experiments run on a shared NVIDIA A40 GPU (44\,GB) under PyTorch with CUDA expandable-segments allocation and Python 3.11.
Each 27K seed takes roughly 48 to 58 minutes; TSV seeds are faster.
The harness picks the least-loaded GPU and retries on transient out-of-memory failures, so results are insensitive to co-tenant load.

\medskip\noindent\textbf{Code availability.} Code, per-seed results, and figure-generation scripts are released at \url{https://github.com/ziyi0227/personalized-modality-weighting-audit}.

\bibliography{refs}

\end{document}